\documentclass[
  aps,
  prl,
  reprint,
  amsmath,
  amssymb
]{revtex4-1}
\setcitestyle{square}
\usepackage{graphicx}

\usepackage{siunitx}
\usepackage{url}

\usepackage{bm}

\newcommand{\figref}[1]{Fig.\,\ref{#1}}

\newcommand{\phiFricJam}{\phi_{\mathrm{J}}^{\mu}}
\newcommand{\phiJam}{\phi_{\mathrm{J}}^{0}}

\newcommand\PeNum{\mbox{Pe}}  

\begin{document}

\title{Discontinuous Shear Thickening in Brownian Suspensions by Dynamic Simulation}

\author{Romain Mari}\altaffiliation{Current address: Department of Applied Mathematics and Theoretical Physics, Centre for Mathematical Sciences, University of Cambridge, Cambridge CB3 0WA, United Kingdom}
\affiliation{Benjamin Levich Institute, %
City College of New York, New York, NY 10031, USA}
\author{Ryohei Seto}
\affiliation{Mathematical Soft Matter Unit, Okinawa Institute of Science and Technology, Onna-son, Okinawa, 904-0495, Japan}
\affiliation{Benjamin Levich Institute, %
City College of New York, New York, NY 10031, USA}
\author{Jeffrey F. Morris}
\affiliation{Benjamin Levich Institute, %
City College of New York, New York, NY 10031, USA}
\affiliation{Department of Chemical Engineering, %
City College of New York, New York, NY 10031, USA}
\author{Morton M. Denn}
\affiliation{Benjamin Levich Institute, %
City College of New York, New York, NY 10031, USA}
\affiliation{Department of Chemical Engineering, %
City College of New York, New York, NY 10031, USA}

\date{\today}




\begin{abstract}
Dynamic particle-scale numerical simulations are used to show that
the shear thickening observed in dense colloidal, or Brownian, suspensions is of a similar nature
to that observed in non-colloidal suspensions,
i.e., a stress-induced transition from a flow of lubricated near-contacting particles to a flow
of a frictionally contacting network of particles.
Abrupt (or discontinuous) shear thickening is found to be a geometric rather than hydrodynamic phenomenon; it stems from the strong sensitivity of the jamming volume fraction to the nature of contact forces between suspended particles.
The thickening obtained in a colloidal suspension of purely hard frictional spheres
is qualitatively similar to experimental observations.
However, the agreement cannot be made quantitative with only hydrodynamics, frictional contacts and Brownian forces.
Therefore the role of a short-range repulsive potential mimicking the stabilization
of actual suspensions on the thickening is studied.
The effects of Brownian and repulsive forces on the onset stress can be combined in an  additive manner.
The simulations including Brownian and stabilizing forces show excellent agreement with experimental data for the viscosity $\eta$ and the second normal stress difference $N_2$.
\end{abstract}

\maketitle

\paragraph{Introduction}

The rheology of dense suspensions is of considerable theoretical and technological importance,
yet the shear rheology of even the simplest case of a suspension of hard spheres
in a Newtonian suspending fluid is incompletely understood~\citep{Stickel_2005}.
Many of the features observed in these suspensions, including shear thinning~\citep{Larson_1999} or thickening~\citep{Barnes_1989,Brown_2014} and the magnitudes and even the algebraic signs of normal stress differences~\citep{Dbouk_2013}, are at best understood at a qualitative level, and a general theoretical framework is lacking.
Furthermore, there has been a tendency to treat the rheology of Brownian (colloidal) suspensions and non-Brownian suspensions as distinct.

Recently, a picture has emerged in which
central aspects of the rheology of non-Brownian dense suspensions are interpreted
as manifestations of proximity to jamming transitions in the parameter space.
These transitions are singularities whose
locations in the volume fraction $\phi$ and shear stress $\sigma$ plane depend on
the details of the microscopic interactions
(shape of the particles, friction, interparticle forces).
In turn, the locations of these singularities shape the large $\phi$ portion of the rheological landscape, i.e., the
effective viscosity and the normal stresses as functions of $\phi$ and $\sigma$.
In particular, in the ``stress-induced friction'' scenario~%
\citep{Brown_2014,Brown_2009,Brown_2012,Fernandez_2013,Seto_2013a,Heussinger_2013,Mari_2014,Mari_2015,Cates_2014},
shear thickening is a transition from a rheological response dominated by
frictionless jamming to one controlled by frictional jamming upon increase of the shear stress.
This transition is argued to be due to the creation of frictional contacts
between particles at high stresses; the contacts are prevented at low stresses
by a short-range stabilizing repulsive force,
as would be expected to be present to stabilize a colloidal dispersion against aggregation by,
for example, attractive van der Waals forces~\citep{Israelachvili_2011,Royall_2013}.
This picture contrasts with previous models for Brownian suspensions~\citep{Brady_1985, Phung_1996, Foss_2000},
in which frictional contacts are neglected based on idealized lubrication hydrodynamics.


Suspensions are said to be colloidal, or Brownian,
when the immersed particles are small enough:
a commonly accepted upper bound for Brownian motion to be significant is a diameter of \SI{1}{\micro\meter}~\citep{Larson_1999}.
For these systems, the Brownian forces have been seen to be an essential
factor in non-Newtonian behavior~\citep{Russel_1992}.
Physically, for a system of strictly hard colloidal spheres under shear in the Stokes regime,
there are only two independent time scales:
the inverse shear rate $\dot\gamma^{-1}$ and the diffusion time $a^2/D_0$;
here $a$ is the sphere radius and $D_0$ is the single-particle diffusion coefficient,
which is related to the suspending fluid viscosity $\eta_0$ and thermal energy $k_{\mathrm{B}}T$
through the Stokes-Einstein relation $D_0 = k_{\mathrm{B}}T/ 6\pi \eta_0 a$.
The shear rate dependence of the rheology can be stated in terms of a competition
between advection and diffusion described by the P\'eclet number
$\PeNum \equiv 6\pi \eta_0 a^3 \dot{\gamma}/kT$.
Smooth spheres with ideal lubrication resulting from hydrodynamic interactions would be non-contacting,
and hence exhibit the following rheology:
a shear rate independent regime close to thermal equilibrium
(that is for $\dot\gamma \lesssim \tau_{\alpha}^{-1}$,
where $\tau_{\alpha}$ is the ``caging'' time,
or the typical time for which the thermal motion leads to a structural reorganization~%
\citep{Cavagna_2009}) where most forces are Brownian,
followed by a shear thinning regime at intermediate values of $\PeNum$,
over which the Brownian forces become progressively less important relative to
the hydrodynamic ones, and finally a shear thickening regime at large $\PeNum$
that is dominated by the hydrodynamic lubrication forces due to increasingly smaller inter-particle
gaps~\citep{Brady_1985, Phung_1996, Foss_2000}.
$\tau_{\alpha}$ diverges at the glass transition $\phi=\phi_{\mathrm{G}}$,
the system develops a yield stress $\sigma_{\mathrm{y}}$ above $\phi_{\mathrm{G}}$,
and the low $\PeNum$ viscosity plateau yields to an asymptotic shear thinning regime
$\eta \sim \sigma_{\mathrm{y}}/\dot{\gamma}$ for $\dot{\gamma} \to 0$~%
\citep{Pusey_1986,Brambilla_2009,Russel_2013}.
Although numerical simulations including only hydrodynamic interactions agree well with experimental data
in the shear thinning regime,
the simulated shear thickening regime is much weaker than is often experimentally observed,
with the disagreement increasing as the volume fraction increases~\citep{Foss_2000}.
%


%
Most of the experimentally studied thickening suspensions are in the colloidal size range~\citep{Brown_2014}.
(Notable exceptions include cornstarch suspensions.)
It is thus essential to address the validity of the stress-induced friction scenario
for these systems.
In this scenario, the number of frictional contacts directly depends on the ratio of the shear stress to the Brownian stress scale $\sigma a^3/k_{\mathrm{B}}T$.
At small stresses, i.e., when $\sigma a^3/k_{\mathrm{B}}T \ll 1$, the thermal motion keeps particles separated and makes contacts unlikely.
(In the equilibrium limit $\mathrm{Pe} \to 0$, the average contact number per particle
is zero at volume fractions below the jamming transition,
otherwise the pressure would diverge as required by the virial equation for hard particles~\citep{Hansen_2006}.)
When $\sigma a^3/k_{\mathrm{B}}T \gg 1$, however, the Brownian forces are not strong enough to overcome
the forces bringing particles together due to the shear, and contacts are created.
For dense suspensions, this shear activated friction mechanism can be related to
a jamming transition framework:
at the largest shear stress for which shear thinning occurs,
the rheology is dominated by the proximity to the frictionless jamming transition point at $\phiJam$,
in the sense that rheological properties are (roughly) diverging functions of the form
${(\phi-\phiJam)}^{-\lambda_0}$,
whereas at shear stresses above shear thickening, the frictional jamming transition at $\phiFricJam$ dominates
and rheological properties scale with ${(\phi-\phiFricJam)}^{-\lambda_{\mu}}$,
with $\lambda_0$ and $\lambda_{\mu}$ two positive exponents (whose exact values are still debated).
Since $\phiFricJam < \phiJam$, this leads to shear thickening,
which can be continuous or discontinuous depending on the proximity to $\phiFricJam$~\citep{Mari_2014,Wyart_2014}.
%


In this work, we show that simulations of frictional colloidal suspensions can reproduce
both continuous and discontinuous shear thickening, hence demonstrating that the stress-induced friction
scenario extends to the Brownian case.
Quantitative agreement with experiments cannot, however, be achieved with only frictional contacts
and Brownian motion combined with hydrodynamic lubrication interactions.
Instead, one must also consider the repulsive force that is induced
between immersed colloids by the stabilization process.
We study the qualitative influence of the range and amplitude of the repulsive force on the shear thickening.
In particular, we show that the effects of the Brownian and stabilizing forces on the onset stress are additive.
For a suitable choice of amplitude and range of the repulsive force,
the simulations of the relative viscosity and the second normal stress difference
(which is large relative to the first) are in good agreement with recent experiments
by Cwalina and Wagner~\citep{Cwalina_2014}.


\paragraph{Model description}

\begin{figure}[t]
  \centering
  \includegraphics[width=0.48\textwidth]{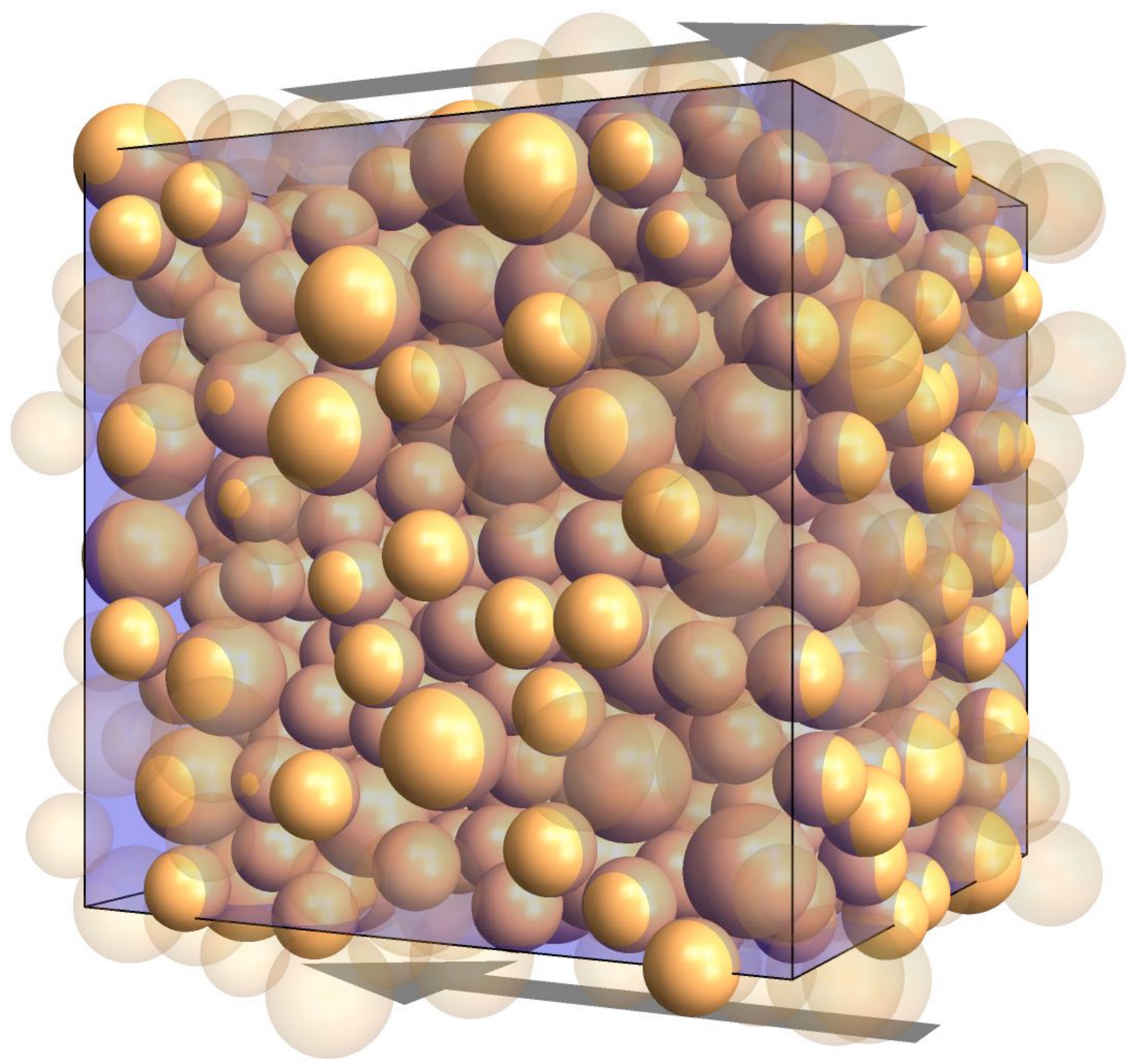}
  \caption{%
    The simulations consider a three-dimensional system of bidisperse spheres under simple shear with Lees-Edwards periodic boundary conditions.}\label{fig:snapshot}
\end{figure}

We study colloidal suspensions of hard spheres at positions $\bm{X}$ and velocities $\bm{U}$
in Stokes flow interacting through hydrodynamic ($\bm{F}_{\mathrm{H}}$), contact ($\bm{F}_{\mathrm{C}}$),
stabilizing repulsive ($\bm{F}_{\mathrm{R}}$), and Brownian ($\bm{F}_{\mathrm{B}}$) forces.
We use a bidisperse system, with spheres of radii $a$ and $1.4a$
at equal volume fractions.
All our results are obtained with $N=500$ particles in a cubic box with Lees-Edwards boundary conditions, as depicted in~\figref{fig:snapshot}.
The equation of motion is given by the following overdamped Langevin equation:
\begin{equation}
  \bm{0}
  =
  \bm{F}_{\mathrm{H}}
  +
  \bm{F}_{\mathrm{C}}
  +
  \bm{F}_{\mathrm{R}}
  +
  \bm{F}_{\mathrm{B}}.
  \label{eq:langevin}
\end{equation}


The hydrodynamic forces are the sum of a drag due to the motion
relative to the surrounding fluid and a resistance to the deformation
imposed by the flow:
$ \bm{F}_{\mathrm{H}}=-\bm{R}_{\mathrm{FU}}(\bm{X}) \cdot
 \bigl(\bm{U}-\bm{U}^{\infty} \bigr) + \bm{R}_{\mathrm{FE}}(\bm{X}):\bm{E}^{\infty} $,
where
$\bm{U}^{\infty}_i = \dot\gamma y_i \bm{\hat{e}}_x$ is the background imposed flow and
$\bm{E}^{\infty} \equiv (\bm{\hat{e}}_x \bm{\hat{e}}_y +\bm{\hat{e}}_y\bm{\hat{e}}_x)\dot\gamma/2$
is the strain rate tensor.
For the dense suspensions of interest in this work,
it is assumed that the hydrodynamic interactions are dominated
by near-contact lubrication interactions~\citep{Ball_1997}
and the long-range interactions are neglected.
For smooth hard spheres, $\bm{R}_{\mathrm{FU}}$ and $\bm{R}_{\mathrm{FE}}$
contain lubrication terms that diverge with vanishing interparticle gap $h$.
We keep only the leading order of these divergences
(these are terms in $\tilde{h}^{-1}$ and $\log \tilde{h}^{-1}$,
with $\tilde{h}=h/a$, for, respectively, the normal and tangential relative motion),
and we regularize these terms with a microscopic cutoff $\delta$
to mimic the particle roughness~\citep{Seto_2013a}.
[That is, the included terms scale as ${(\tilde{h}+\delta/a)}^{-1}$ and $\log{(\tilde{h}+\delta/a)}^{-1}$.]
The results presented here are obtained with $\delta=10^{-3}a$.
The results depend only weakly on the value of $\delta$.
The listing of the individual matrix elements is given in~\citep{Mari_2014}.
%


Contacts are modeled by a linear spring consisting of both normal and tangential components,
which is a simple model commonly used in granular physics~\citep{Mari_2014}.
Tangential and normal components of the contact force
$\bm{F}_{\mathrm{C}}^{(i,j)}$ between two particles satisfy Coulomb's friction law
$\bigl|\bm{F}_{\mathrm{C,tan}}^{(i,j)} \bigr| \leq \mu \bigl|\bm{F}_{\mathrm{C,nor}}^{(i,j)}\bigr|$.
The spring constants are chosen for each $\PeNum$ and $\phi$ so that the average minimal center-to-center distance $d_{ij}$ between any two particles $i$ and $j$ is maintained as $1-d_{ij}/(a_i+a_j)\approx 0.02$ (here $a_i$ and $a_j$ are the particle radii).


We take a stabilizing repulsive force that decays exponentially with the
interparticle gap $h$ as $|\bm{F}_{\mathrm{R}}|= F^{\ast} \exp(-h/\lambda)$,
with a characteristic length $ \lambda$.
This provides a simple model of screened electrostatic
interactions that can often be found in aqueous systems~\citep{Laun_1994,Franks_2000}, in which case $\lambda$ is the Debye length.
With one force scale and one length scale, this
is the most basic parametrization of a generic stabilizing force.

\begin{figure}[t]
  \centering
  \includegraphics[width=0.48\textwidth]{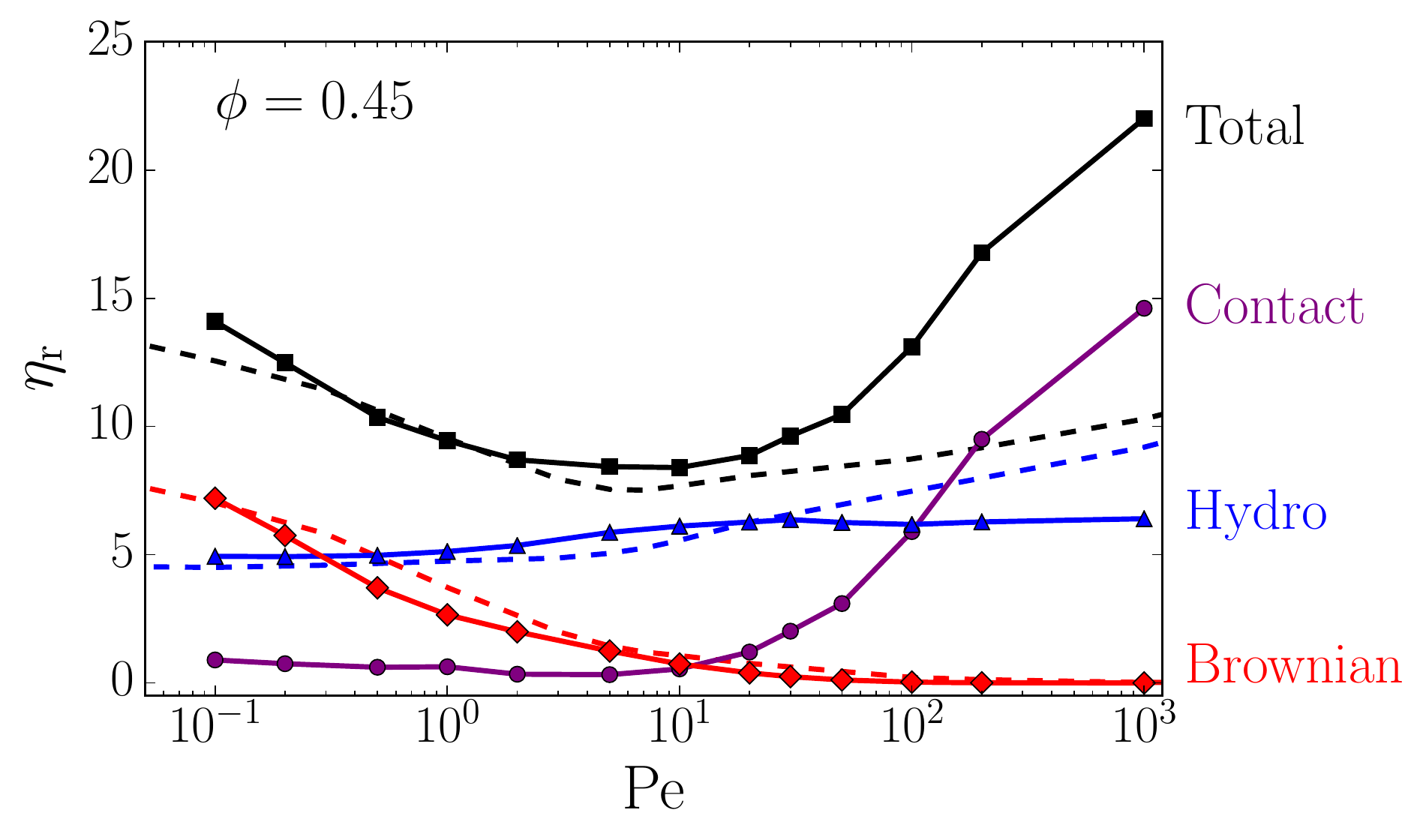}
  \caption{%
    Comparison of our simulation results (solid lines) for the viscosity
    as a function of $\PeNum$ with results obtained with
    Stokesian Dynamics by Foss and Brady~\citep{Foss_2000} (dashed lines)
    for a purely Brownian suspension at $\phi=0.45$.
    We plot the total relative viscosity (black squares) as well as the individual hydrodynamic
    (blue triangles), Brownian (red diamonds) and contact (purple circles) contributions.}\label{fig:comparison_SD}
\end{figure}


The Brownian forces acting on different particles are correlated through
the hydrodynamic interactions,
and their second order cumulant is given by the fluctuation-dissipation theorem~\citep{Deutch_1971}.
In a simulation with discretized time $t_n$ with intervals $\Delta t$,
this becomes~\citep{Ermak_1978,Bossis_1987}
\begin{equation}
  \bigl\langle
    \bm{F}_{\mathrm{B}}(t_m)
    \bm{F}_{\mathrm{B}}(t_n)
  \bigr\rangle - \langle \bm{F}_{\mathrm{B}}(t_m) \rangle \langle \bm{F}_{\mathrm{B}}(t_n) \rangle
  = \frac{2 k_{\mathrm{B}}T}{\Delta t} \bm{R}_{\mathrm{FU}}(\bm{X}) \delta_{mn}
  \label{eq:brownian_force_dev}
\end{equation}
Owing to the dependence of the Brownian force term on the configuration
as shown by Eq.\,\eqref{eq:brownian_force_dev},
the Langevin equation~\eqref{eq:langevin} contains a multiplicative noise.
As a consequence, we must specify by which convention we evaluate the Brownian force~\citep{Kampen_1992}.
A natural choice for numerical simulation is the It\=o convention,
which we adopt here; that is, when $m=n$ in Eq.\,\eqref{eq:brownian_force_dev},
we evaluate $\bm{R}_{\mathrm{FU}}$ at the beginning of the time step
as $\bm{R}_{\mathrm{FU}}(\bm{X}(t_n))$.
In this convention, the Brownian forces have a non-zero average (called ``drift'')%
\footnote{One can find the same result, starting from the underdamped Langevin equation,
by integrating the equation over a time step $\Delta t$ much larger than
the inertial time scale $\tau_{\mathrm{d}}=m/(\eta_0 a)$
(where $m$ is the characteristic mass of a colloid)
but much smaller than $a^2/D$ or $\dot\gamma^{-1}$ and then taking the
$\tau_{\mathrm{d}}\to 0$ limit~\citep{Ermak_1978,Brady_1988,Grassia_1995}.}~\citep{Lau_2007}:
\begin{equation}
  \bigl\langle
    \bm{F}_{\mathrm{B}}(t_n)
    \bigr\rangle
  = k_{\mathrm{B}}T \bm{R}_{\mathrm{FU}}(\bm{X}(t_n))\cdot \nabla \cdot \bm{R}_{\mathrm{FU}}^{-1}(\bm{X}(t_n)).
  \label{eq:brownian_force_avg}
\end{equation}


At each time step, we evaluate the contact and repulsive forces $\bm{F}_{\mathrm{C}}$ and $\bm{F}_{\mathrm{R}}$,
generate Brownian forces $\bm{F}_{\mathrm{B}}$
according to Eqs.\,\eqref{eq:brownian_force_dev} and~\eqref{eq:brownian_force_avg}
(through the algorithm of~\citep{Ball_1997}), and we solve Eq.\,\eqref{eq:langevin} for the velocities:
\begin{equation}
  \bm{U}-\bm{U}^{\infty}
  =
  \bm{R}_{\mathrm{FU}}^{-1}
  \cdot
  \left(
    \bm{R}_{\mathrm{FE}}:
    \bm{E}^{\infty}
    +
    \bm{F}_{\mathrm{C}}
    +
    \bm{F}_{\mathrm{R}}
    +
    \bm{F}_{\mathrm{B}}
  \right).
  \label{eq:velo_solution}
\end{equation}



\begin{figure*}[tb]
  \centering
  \includegraphics[width=0.48\textwidth]{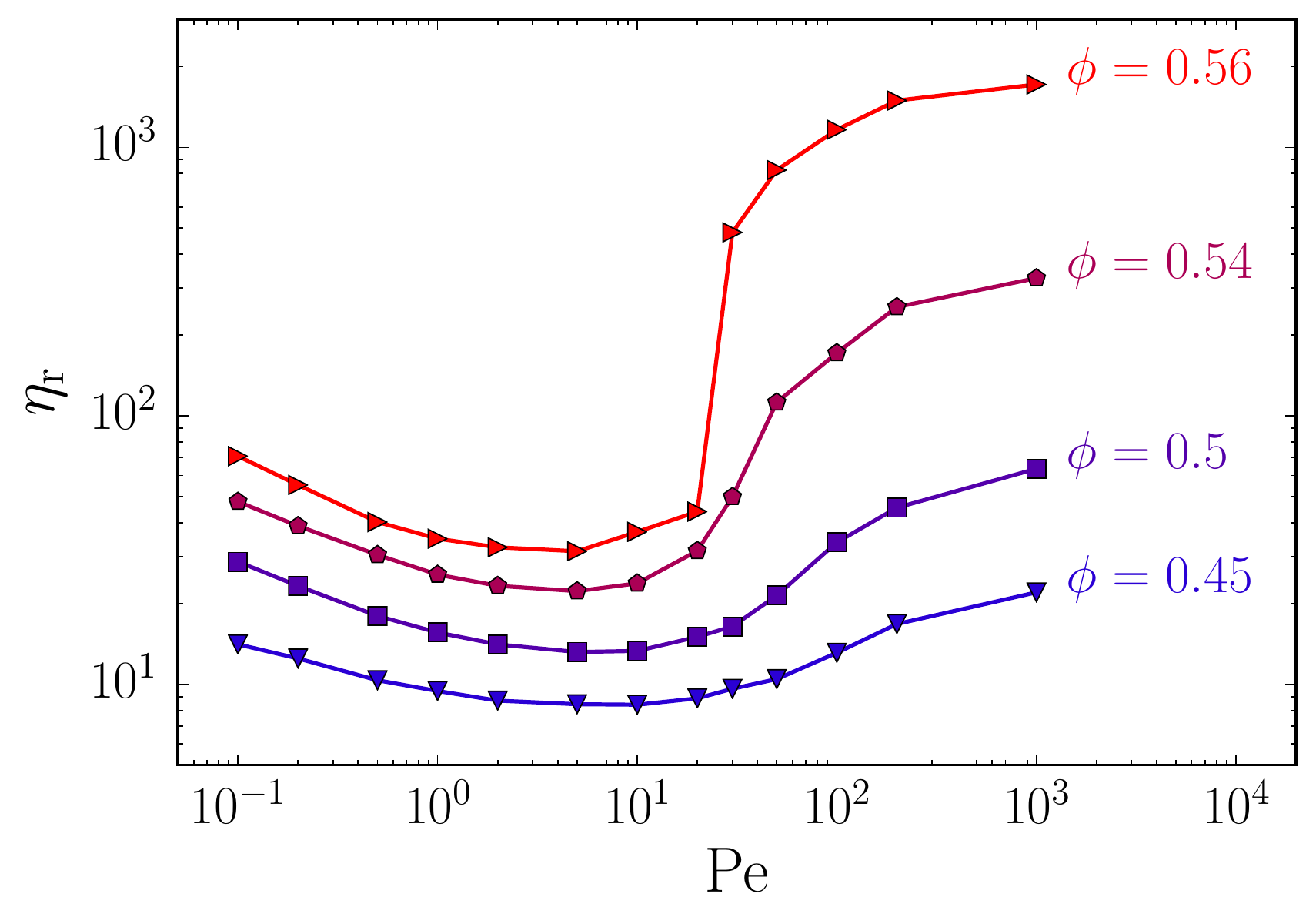}
  \includegraphics[width=0.48\textwidth]{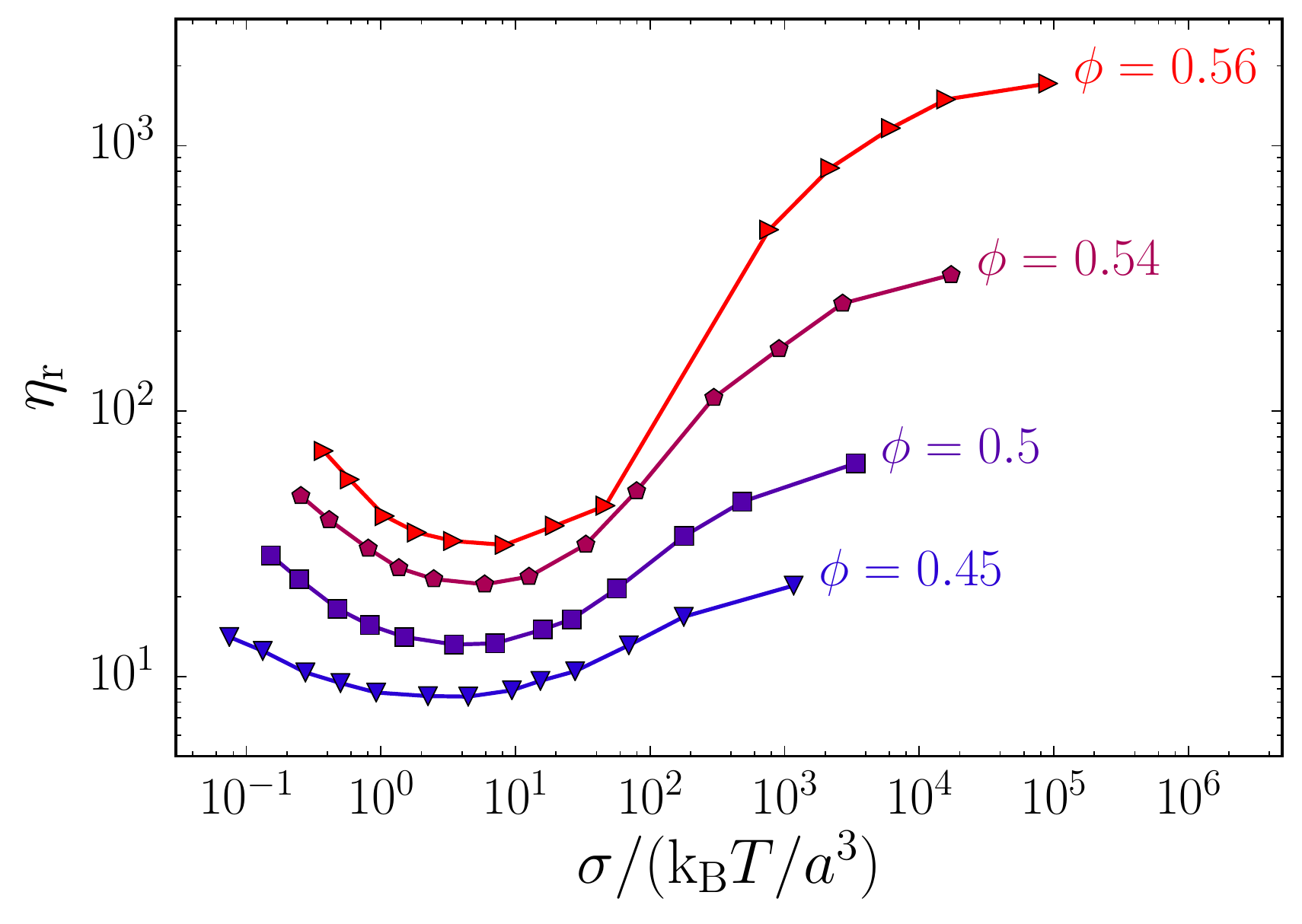}
  \caption{%
    Relative viscosity $\eta_{\mathrm{r}}$ as a function of
    the P\'eclet number $\PeNum$ (left) and as a function of the reduced shear stress (right)
    for a Brownian suspension of hard
    frictional spheres at several volume fractions $0.45\leq \phi \leq 0.56$.
  }\label{fig:rheology_pure_Brownian}
\end{figure*}

\begin{figure*}
  \centering
  \includegraphics[width=0.9\textwidth]{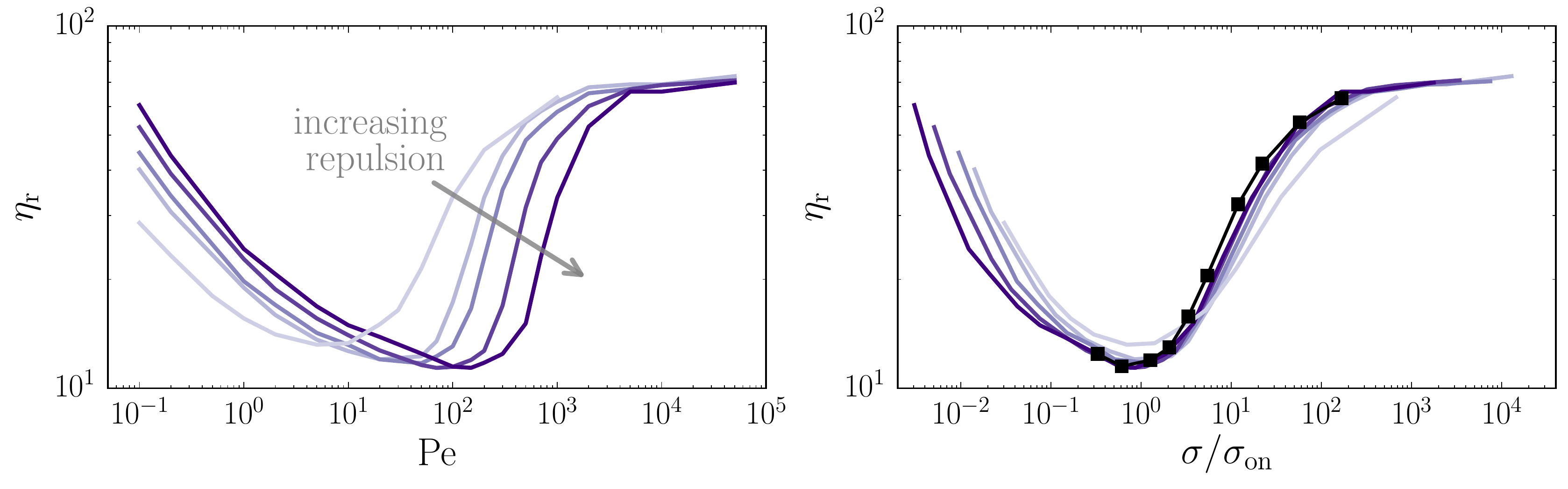}
  \caption{
    \textbf{Left:} The effect of the repulsion amplitude $F^{\ast}$ on the rheology
    of a colloidal suspension for $\phi=0.5$,
    for several values
    $F^{\ast}=0, 10^3 k_{\mathrm{B}} T/a, 2\times 10^3 k_{\mathrm{B}} T/a, 5\times 10^3 k_{\mathrm{B}} T/a$ and $10^4 k_{\mathrm{B}} T/a$ from light to dark color.
    The value of $F^{\ast}$ essentially influences the onset stress of thickening:
    increasing $F^{\ast}$ delays the thickening.
    As $F^{\ast}$ increases, the shear thinning regime is extended and the minimum value
    of the viscosity decreases slightly.
    \textbf{Right:}
    The effect of both the Brownian forces
    and the repulsive force can be summed up by plotting the relative viscosity as a function of the stress rescaled by the onset stress $\sigma_{\mathrm{on}}=5k_{\mathrm{B}} T/a^3+0.01 F^{\ast}/a^2$.
    This collapse of the data extends to the non-Brownian purely repulsive case
    (in black squares),
    for which $\sigma_{\mathrm{on}} = 0.01 F^{\ast}/a^2$.}\label{fig:stress_rescaling}
\end{figure*}

\paragraph{Results}

We first show example results obtained for the ``pure'' Brownian case
without repulsive forces, i.e., $F^{\ast}=0$.
We show in~\figref{fig:comparison_SD} that
the frictional contacts significantly modify the high $\PeNum$ dynamics.
In this figure we compare the relative viscosity $\eta_{\mathrm{r}}$
at a volume fraction $\phi=0.45$ obtained with our simulation using a friction
coefficient $\mu=1$ with the result obtained using Stokesian Dynamics
(with particles not making contact) by Foss and Brady~\citep{Foss_2000}
as a function of $\PeNum$.
There is very good agreement in the shear thinning regime at low $\PeNum$,
where hydrodynamics and Brownian forces are dominant.
At high $\PeNum$, however, where Stokesian Dynamics shows a very mild increase of the viscosity,
our simulation shows a significant thickening that is due to contact stresses
arising from the building up of a contact network.
The hydrodynamic contribution in our simulation does not thicken at high $\PeNum$
as predicted theoretically~\citep{Brady_1997} for hydrodynamically-interacting hard spheres
due to the lubrication cutoff, and instead becomes almost shear rate independent.

\begin{figure}
  \centering
  \includegraphics[width=0.48\textwidth]{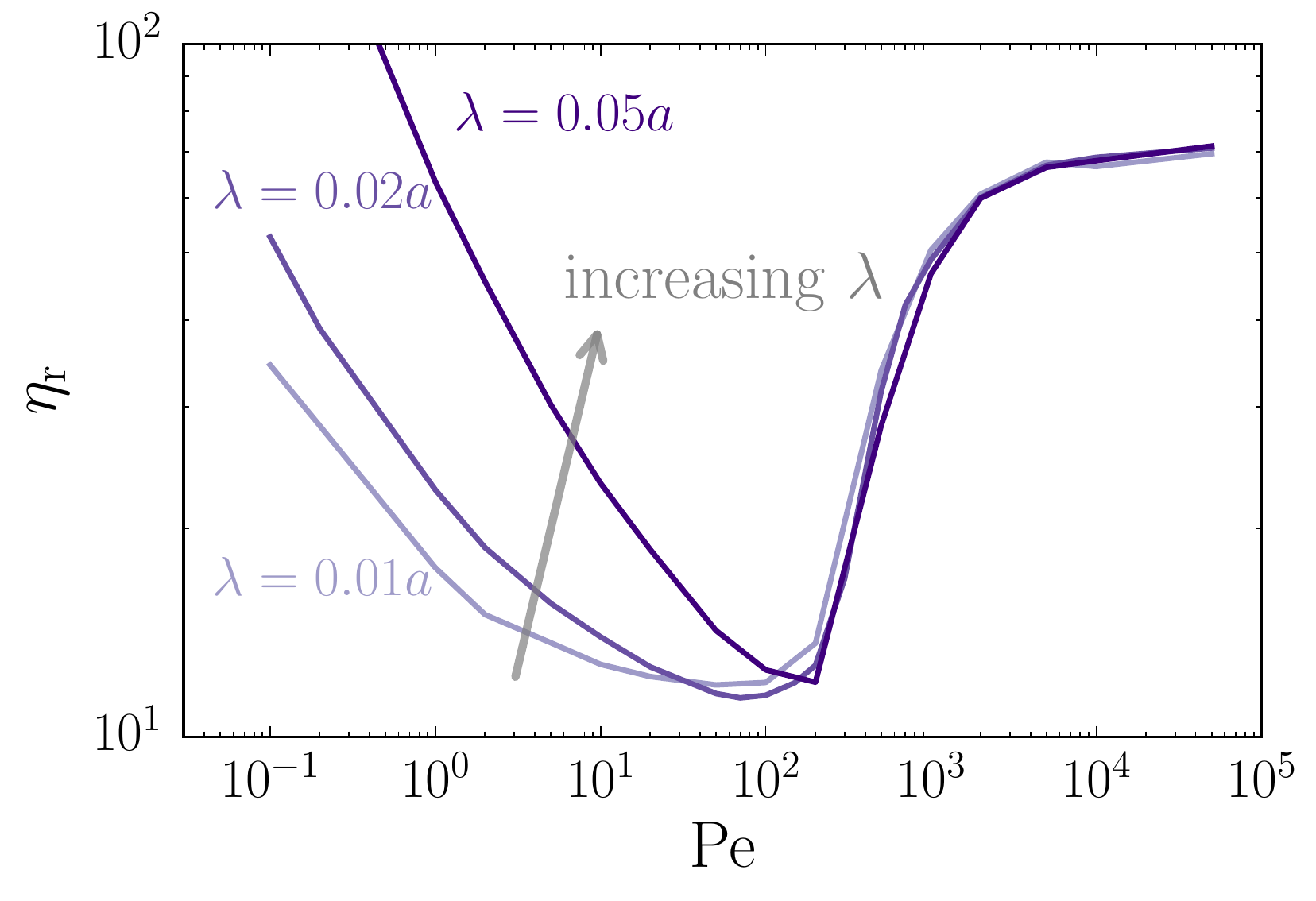}
  \caption{%
    The effect of the repulsion range $\lambda$ on the rheology of a colloidal suspension for $\phi=0.5$,
    from $\lambda/a=0.01$ to $\lambda/a=0.05$, from light to dark color.
    When $\lambda$ increases, the low shear rate viscosity increases,
    and as a consequence the shear thinning regime becomes more pronounced.
    The high shear rate rheology is unaffected by a change of $\lambda$.}\label{fig:comparison_Debye_length}
\end{figure}


The volume fraction dependence of the thickening due to the frictional
contacts is shown in~\figref{fig:rheology_pure_Brownian}.
We are able to obtain both continuous shear thickening for $\phi\lesssim 0.55$
and discontinuous shear thickening for the highest volume fractions $\phi > 0.55$.
The thickening is seen above an onset stress $\sigma_{\mathrm{on}} \approx 5 k_{\mathrm{B}}T/a^3$.
This shows that the shear-induced friction mechanism introduced to explain the thickening of
non-colloidal suspensions~\citep{Fernandez_2013,Seto_2013a,Heussinger_2013,Mari_2014,Cates_2014}
extends to the colloidal case, giving the same qualitative rheology.
The values of the onset stress obtained experimentally are typically of order
$\sigma_{\mathrm{on}} \approx 100 k_{\mathrm{B}}T/a^3$,
however~\citep{Laun_1984,Egres_2005a,Cwalina_2014},
which is substantially larger than the value of about $5 k_{\mathrm{B}}T/a^3$
found in our $F^{\ast}=0$ simulation.
This can be understood as our simulation missing a repulsive force that arises
from the suspension ``stabilization'' in the experiments.
%


We first look at the effect of the repulsive force on the thickening of the colloidal suspension.
The relative viscosity curves for different values of $F^{\ast}$ are shown
in the left panel of \figref{fig:stress_rescaling} for $\phi=0.5$.
The main effect of the repulsion is, as expected, to push the onset of thickening to higher stresses.
The relative viscosity in the thickened state is unaffected by the value of $F^{\ast}$,
as in this regime the repulsive force can be neglected relative to the hydrodynamic and contact forces.
Note that the slope of the shear thinning is also the same for all of the simulated $F^{\ast}$.
In the right panel of \figref{fig:stress_rescaling}, we show that the onset stress
is approximately $\sigma_{\mathrm{on}} \approx 5k_{\mathrm{B}} T/a^3+0.01 F^{\ast}/a^2$.
Thus, to a good approximation the effects of Brownian and repulsive forces on shear thickening
can be combined in a simple additive manner.
In this regard, the Brownian forces have an effect that is virtually identical to that
of a potential repulsive force.


Besides $F^{\ast}$, our repulsive force contains another free parameter, the force decay length $\lambda$.
As we show in~\figref{fig:comparison_Debye_length},
$\lambda$ essentially controls (in conjunction with the Brownian motion) the strength of
the shear thinning at low $\PeNum$.
Increasing $\lambda$, i.e., increasing the distance over which the repulsive force decays,
makes the shear thinning more pronounced.
This can be qualitatively understood: when $\lambda \to 0$, the repulsive force disappears,
and only the shear thinning due to the Brownian motion remains.


\begin{figure}
  \centering
  \includegraphics[width=0.48\textwidth]{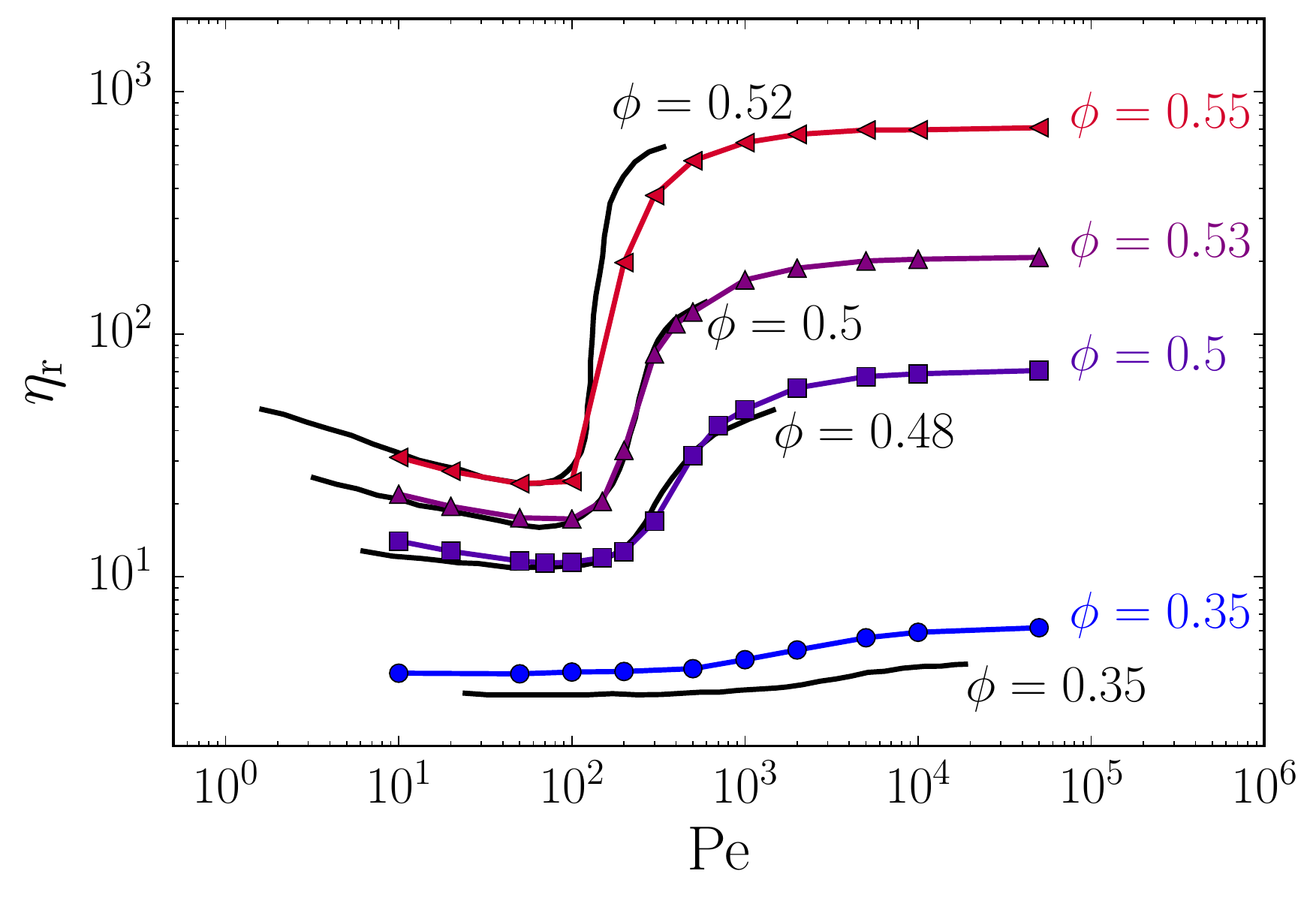}
  \includegraphics[width=0.48\textwidth]{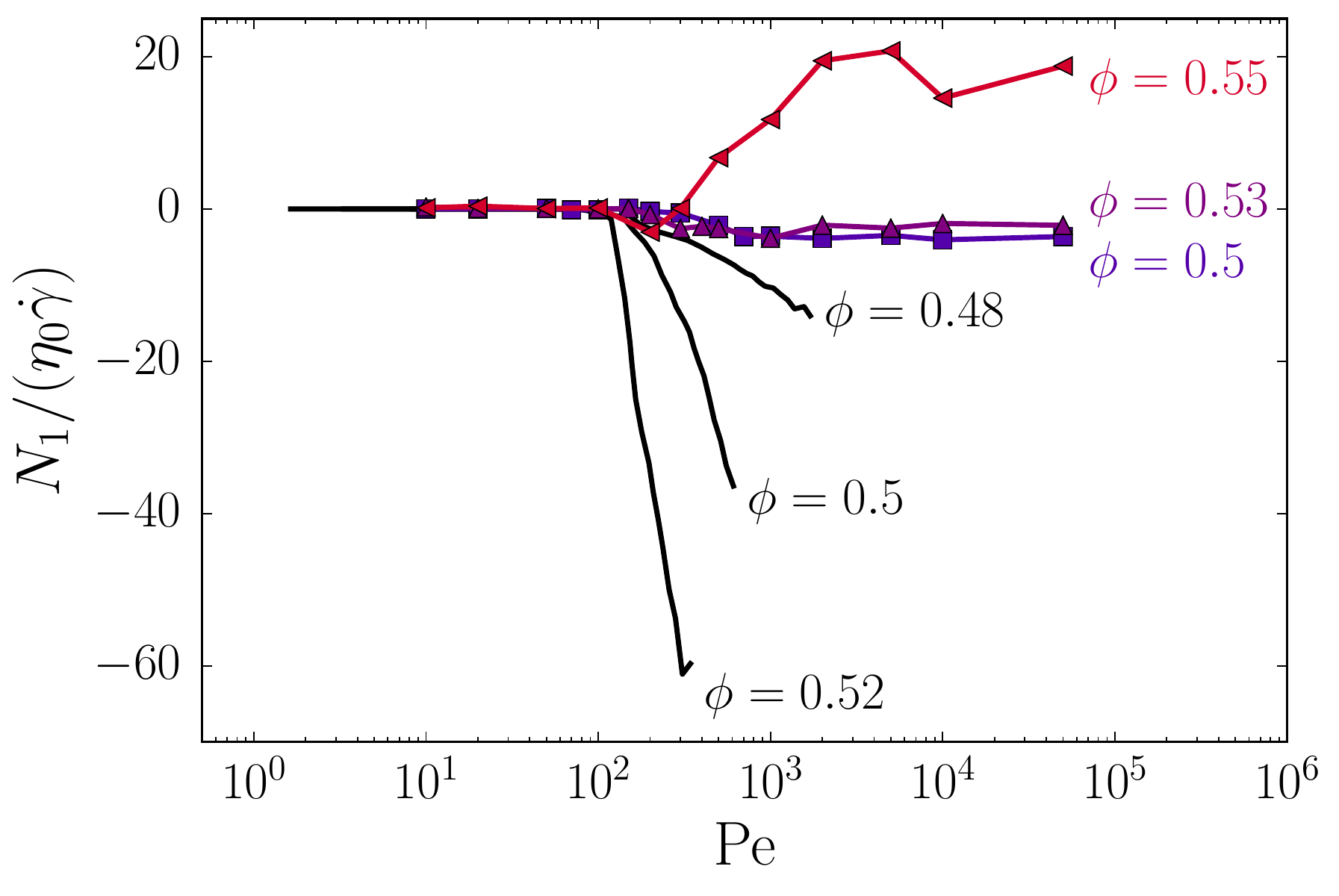}
  \includegraphics[width=0.48\textwidth]{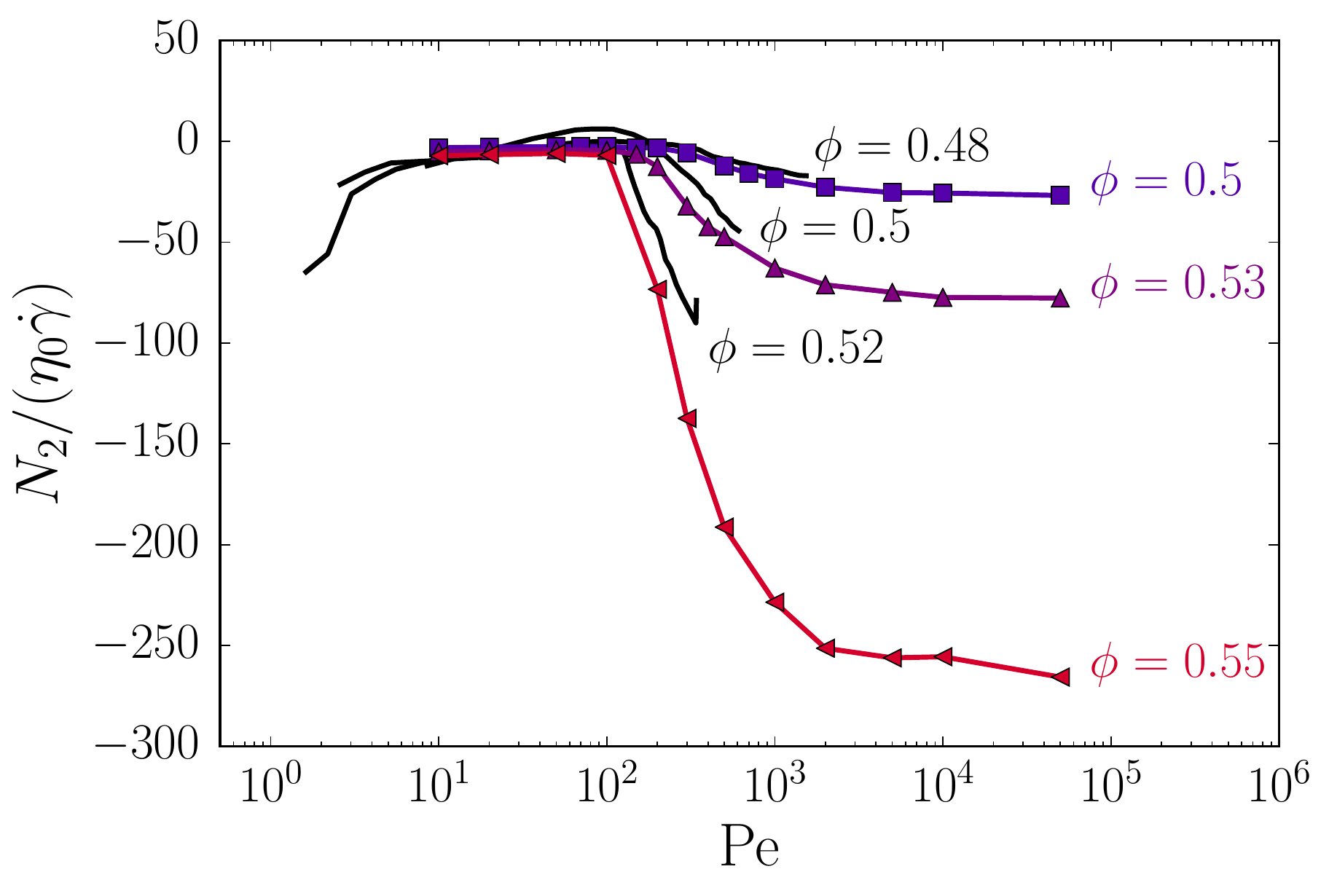}
  \caption{%
    Comparison with experimental data from Cwalina and Wagner~\citep{Cwalina_2014} (black lines)
    for the relative shear viscosity (top),
    second (center) and first (bottom) normal stress difference viscosities
    as functions of the P\'eclet number.
    Simulation results (colored lines and symbols) are obtained
    with a repulsive force at contact $F^{\ast}=5\times 10^3 \mathrm{k_B}T/a$ and
    a repulsion range $\lambda = 0.02a$.}\label{fig:comparison_Cwalina_visc}
\end{figure}

\begin{figure}
  \centering
  \includegraphics[width=0.48\textwidth]{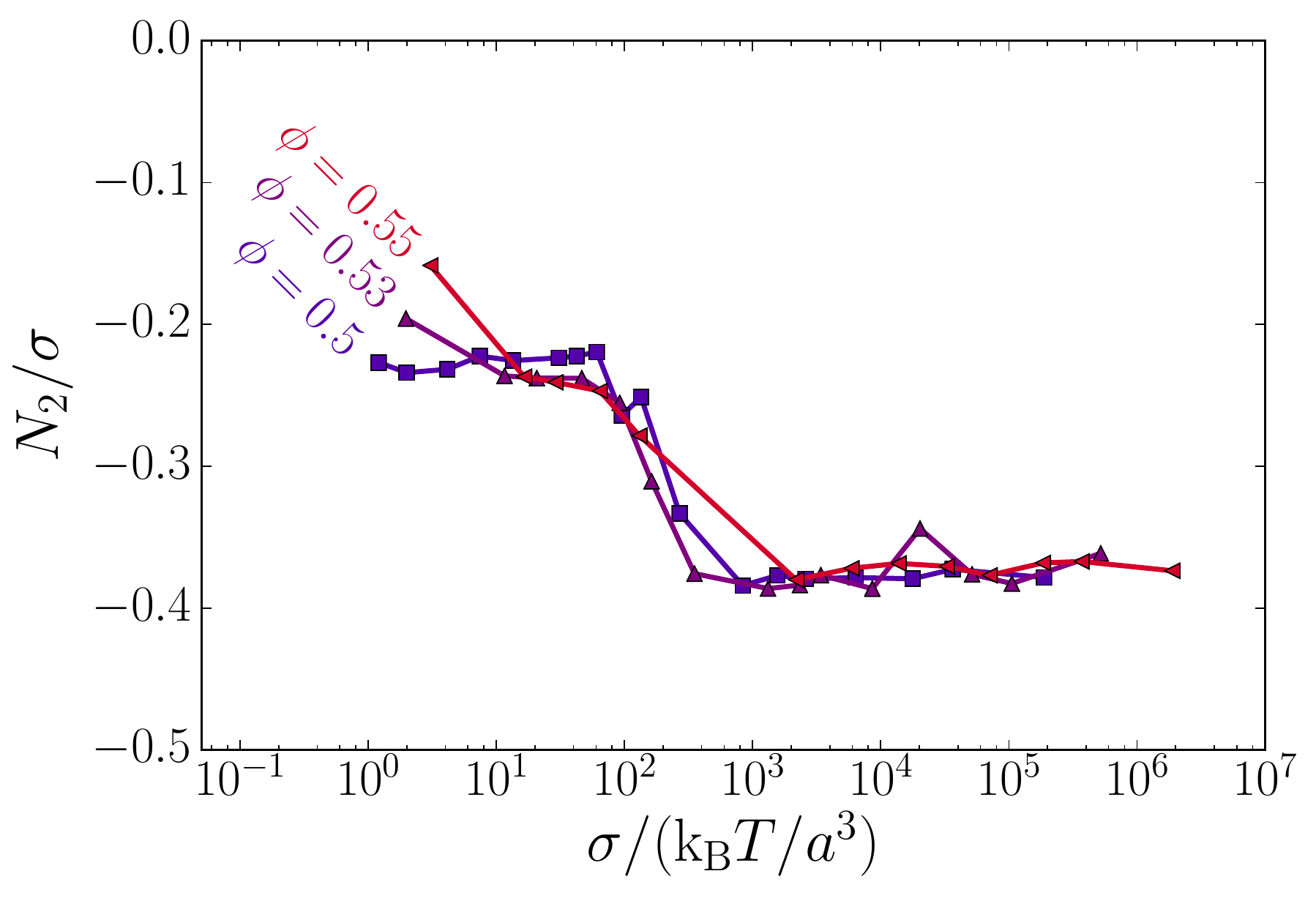}
  \caption{%
    Second normal stress difference $N_2$ normalized by the shear stress as a function of
    the shear stress for several volume fractions $\phi$ for the same conditions
    as in~\figref{fig:comparison_Cwalina_visc}.}\label{fig:N2_over_stress}
\end{figure}

We use simulations to assess the appropriate repulsive force to capture the behavior seen in experiment.
We focus here on the recent data by Cwalina and Wagner~\citep{Cwalina_2014},
which include measurements of the shear stress and normal stress differences for
a suspension of silica beads with radius $a = \SI{260}{\nano\meter}$
in a low molecular weight polyethylene glycol (PEG) Newtonian suspending fluid at $T=\SI{300}{\kelvin}$.
The particles are coated with octadecane chains to provide steric stabilization~\citep{Kalman_2010}.
The short range van der Waals attraction is also reduced by index matching
between particles and solvent.

We obtained the best comparison with the results of Cwalina and Wagner~\citep{Cwalina_2014}
by setting $\mu=1$, $F^{\ast}=5\times 10^3 \mathrm{k_B}T/a$ and $\lambda = 0.02a$,
as shown in \figref{fig:comparison_Cwalina_visc}.
The agreement with the experimental data for the relative viscosity is excellent.
The second normal stress difference $N_2$ also shows a very good agreement with the experimental data,
being negative for all volume fractions; $-N_2/\sigma$ is
in the range $0.15$--$0.4$ for all $\PeNum$, as shown in~\figref{fig:N2_over_stress},
consistent with the behavior for non-Brownian suspensions~\citep{Mari_2014}.
Rather surprisingly, given the agreement for both $\eta_{\mathrm{r}}$ and $N_2$,
$N_1$ disagrees substantively between our simulations and these experiments.
In the experiments, $N_1<0$, as predicted based on hydrodynamic force dominance~\citep{Morris_1999}.
Our simulations find weaker negative $N_1$ for the lower volume fractions presented
($\phi=0.50$ and $0.53$), while for $\phi=0.55$, $N_1>0$.
We note that Lootens~\textit{et al.}\,\citep{Lootens_2005} observed a similar change in sign of $N_1$
at the shear thickening transition.
The parameters used in the simulations can be translated into SI units using the experimental parameters,
with an inferred repulsive force at contact of $F^{\ast} \approx \SI{79}{\pico\newton}$ and
a repulsion range of $\lambda \approx \SI{5.2}{\nano\meter}$,
which is to be compared to the thickness of the stabilizing polymer comb estimated to be
\SIrange{15}{20}{\nano\meter} from the structure factor measured by neutron scattering~\citep{Kalman_2010}.
The volume fractions used in the simulations to get the best agreement
with experiments are always higher than the experimental ones.
This might be attributed to the higher polydispersity used in the simulations,
which lowers the viscosity at a given volume fraction.
%



\paragraph{Discussion}


Our simulation results suggest that the shear thickening of a dense suspension
is fundamentally the same phenomenon
for particles from $\mathcal{O}(\SI{10}{\nano\meter})$ to $\mathcal{O}(\SI{100}{\micro\meter})$.
Abrupt or discontinuous shear thickening
occurs in suspensions close to jamming; as such, it is a geometric
rather than a hydrodynamic phenomenon.
(Hydrodynamics has been shown to
provide a basis for weak continuous shear thickening in dilute~\citep{Brady_1997}
as well as concentrated~\citep{Phung_1996,Foss_2000} suspensions.)
It depends crucially on the existence of a mechanism preventing contacts at small stresses,
such as Brownian motion in a purely colloidal suspension.
Interestingly, for understanding the qualitative rheological behavior,
the Brownian force can be thought of as a potential repulsive force.
A comparison between our numerical results and experimental data~\cite{Cwalina_2014} shows that
in actual suspensions the repulsive effect of the Brownian force adds to an actual potential repulsive force.
This analogy between Brownian force and repulsive force is not restricted to the thickening regime
and has been proposed for suspensions at equilibrium~\citep{Brito_2006}
and, recently, in the shear thinning regime~\citep{Trulsson_2015}.
Our work supports the view that a
theoretical modeling of shear thickening should be centered on
a framework common to Brownian and non-Brownian systems~\citep{Guy_2015},
e.g., a geometric description~\citep{Lerner_2012,Degiuli_2015} including
a stress-induced friction mechanism.

\section*{ACKNOWLEDGMENTS}
Our code makes use of the CHOLMOD library by Tim Davis
(\url{https://www.cise.ufl.edu/research/sparse/cholmod/}) for direct
Cholesky factorization of the sparse resistance matrix.
This research was supported in part by a grant of computer time from the City
University of New York High Performance Computing Center under NSF
Grants CNS-0855217, CNS-0958379, and ACI-1126113.
J. F. M. was supported in part by NSF PREM (DMR 0934206).

\bibliography{dst}

\end{document}